\documentstyle[11pt]{article} 
\def\v1{\vspace{1cm}} 
\def\be{\begin{equation}} 
\def\ee{\end{equation}} 
\def\bc{\begin{center}} 
\def\ec{\end{center}} 
\def\ik{\partial} 
\def\vh{\varphi}

\newcommand{\bea}{\begin{eqnarray}} 
\newcommand{\eea}{\end{eqnarray}}

\begin{document} 
 
\title 
{\bf Conformal Unification of General Relativity and Standard Model} 
 
\author{ 
M. Pawlowski\\ 
{\normalsize\it Soltan Institute for Nuclear Studies} \\ 
{\normalsize\it Warsaw,Poland.}\\[0.3cm] 
V.V. Papoyan\\ 
{\normalsize\it Yerevan State University, 375049, Yerevan, Armenia} \\[0.3cm] 
V.N. Pervushin, V.I. Smirichinski \\ 
{\normalsize\it Joint Institute for Nuclear Research},\\ 
 {\normalsize\it 141980, Dubna, Russia.} 
} 
\date{\empty} 
\maketitle 
 
 
\begin{abstract} 
{\large 
{ 
The unification of general relativity and standard model for strong and 
electro-weak interactions is considered on the base of the conformal symmetry 
principle. The Penrose-Chernikov-Tagirov Lagrangian is used 
to describe the Higgs scalar field modulus and gravitation. We show 
that the procedure of the Hamiltonian reduction converts the homogeneous part 
of the Higgs field into the dynamical parameter of evolution 
of the equivalent reduced system. The equation 
of dynamics of the ``proper time'' of an observer with respect to 
the evolution parameter reproduces the Friedmann-like 
equation, which reflects the cosmological 
evolution of elementary particle masses. The value of the Higgs field 
is determined, at the present time, 
by the values of mean density of matter and 
the Hubble parameter in satisfactory agreement with the data of 
cosmological observations.}}

\end{abstract} 
 
 
 
\section{Introduction.}

The Standard Model (SM) for electroweak and strong 
interactions is almost established for phenomena 
up to 100GeV; one only needs to observe the Higgs particle 
in experiment and manage to include gravity into the unified theory. 
The conventional scheme \cite{1} of 
the minimal coupling of the scalar field with gravity 
supposes naive adding of the General Relativity (GR) 
and SM, each of these having own dimensional parameters. 
In this scheme, there is a number of 
difficulties connected with the existence of a scalar 
mode of a nonvanishing vacuum expectation value in 
cosmology \cite{1}.

Another more fundamental way of the unification of GR and SM is to propose 
that there is only one universal dimensional parameter 
for all interactions and all regions of energies. 
First attempt to describe the Newton coupling constant in GR as the 
vacuum averaging of the same Higgs field were made in 1974 
 \cite{1974} with the idea that spontaneous 
symmetry breaking forms simultaneously the scale of masses in both GR and SM. 
(In the context of the change of the Newton coupling constant in GR 
by the scalar field, we should also recall the Jordan-Brans-Dicke 
scalar tensor theory ~\cite{jbd}.) 
 
The next essential step on the way of decreasing the number  of dimensional 
parameters in the Lagrangian of the unified theory of GR and SM was made in 
paper~\cite{PR}, where  the conformal 
invariant unified theory was considered without any dimensional parameters. 
The theory represents 
the standard model of strong and 
electro-weak interactions in which gravitation and the modulus of the 
Higgs scalar field 
are described by the Penrose-Chernikov-Tagirov Lagrangian~\cite{PCT}, 
which has no any dimensional parameters. 
 
In the present paper, we investigate the dynamics of a scalar field  in 
the conformal unified theory (CUT) GR and SM \cite{PR,psp}. 
We use the Dirac-ADM parametrization of the metric~\cite{ADM,Y,K} and 
the Lichnerowicz conformal invariant variables ~\cite{AL} 
constructed with the help of the space scale 
component of the initial metric.

\section{Conformal Unification of GR and SM}

The action of the conformal invariant theory of GR and SM is a sum 
of these theories 
\be 
\label{model} 
{W}_{CUT}= W_{PCT}+W_{SM}^c, 
\ee 
where the Penrose-Chernikov-Tagirov (PCT) action $W_{PCT}$ 
\be 
\label{pct} 
W_{PCT}(\vh_{_{PCT}},g)= \int d^4x[-\sqrt{-g}R(g){\vh_{_{PCT}}^2 \over 6} 
+\vh_{_{PCT}}\partial_\mu(\sqrt{-g}g^{\mu\nu}\partial_\nu\vh_{_{PCT}})] 
\ee 
describes the metric and the PCT scalar field $\vh_{_{PCT}}$, and 
\be 
\label{smc} 
W_{SM}^c[\vh_{H},{\bf n},V,\psi,g] =\int d^4x\left({\cal 
L}_0^{SM}+\sqrt{-g}[-\vh_{H}F+\vh_{H}^2B-\lambda\vh_{H}^4]\right) 
\ee 
is the conformally invariant part of the SM action (i.e. the conventional SM 
action without the ``free" part for the modulus of the Higgs $SU(2)$ doublet 
$\vh_H$ and without the Higgs mass term), 
$B$ and $F$ are the mass terms of the vector $V$ and fermion $\psi $ fields, 
respectively 
\be 
\label{66} 
B=D{\bf n}(D{\bf n})^*\,;\,F=(\bar\psi_L{\bf n})\psi_R+ h. c.;~~~~ 
{\bf n}=\left(\begin{array}{c} {\rm n}_1 \\ 
{\rm n}_2 \end{array}\right);\;\; 
{\rm n}_1\stackrel{*}{{\rm n}_1}+n_2\stackrel{*}{{\rm n}_2}=1, 
\ee 
${\bf n}$ is the angular component of the Higgs $SU(2)$ doublet 
 
The conformal symmetry of the Lagrangian (\ref{model}) means 
that it is invariant with respect to simultaneous transformations 
of all fields in the theory, according to the rule 
\be \label{con} 
{}^{(n)} f^{\prime}(x)= {}^{(n)} f(x) \Omega^{n}(x), 
\ee 
where $(n)$ is the conformal weight of the field 
$f$. 
 
The main idea of the present paper (introduced in \cite{PR} and developed in 
\cite{psp}) is to identify the PCT scalar field with the modulus of the Higgs 
doublet within the rescaling factor $\chi$ 
\be \label{67a} 
\vh_H=\chi\vh_{_{PCT}}. 
\ee 
The rescaling factor $\chi$ must be regarded as a new coupling constant, 
which coordinates weak and gravitational scales \cite{PR}. The value of  
rescaling factor is not predicted by the present theory and must be given by 
experiment. 
 
\section{Hamiltonian and Evolution Parameter} 
 
The Hamiltonian description, 
in general relativity, is achieved by the $(3+1)$ 
foliation of the four-dimensional manifold \cite{ADM} 
\be 
\label{28} 
  (ds)^2=g_{\mu\nu}dx^\mu dx^\nu= N^2 dt^2-{}^{(3)}g_{ij}\breve{dx}{}^i 
  \breve{dx}{}^j\;;\;\;(\breve{dx}{}^i=dx^i+N^idt) 
\ee 
Our model differs from the conventional Einstein theory 
by an additional local conformal symmetry 
(\ref{con}). 
This symmetry allows us to eliminate one degree of 
freedom which is formally present in the Lagrangian but for which there is 
no dynamical equation of motion.  We can choose the retransformation field 
parameter $\Omega=\Omega_c$ in such a way that the space scale factor 
$||{}^{(3)}g||$ is eliminated from the observable variables and the interval 
\be 
\label{sfix} ||{}^{(3)}g_c||=\Omega_c^6||{}^{(3)}g||=1; 
~~~~~~~~~~~~ (ds)_c^2= 
N_c^2 dt^2-{}^{(3)}g_{(c)ij}\breve{dx}{}^i \breve{dx}{}^j. 
\ee 
Then, the space 
volume $\int d^3x \sqrt{{}^{(3)}g_c}=\int d^3x $ becomes an integral of 
motion. The new metric $N_c, {}^{(3)}g_c $ and new variables  
\be  
\label{9} 
{}^{(n)} 
f_c(x)= {}^{(n)} f(x) \Omega^{n}_{c}(x);\;\;\;  
\vh_{H_c}=\vh_H\Omega_c;\;\;\; \vh_c\buildrel{\rm 
def}\over 
=(\vh_{PCT})_c=\vh_{PCT}\Omega_c 
\ee 
coincide with the conformal variables 
introduced in GR by Lichnerowicz \cite{AL}, which are very convenient for 
studying the problem of initial data \cite{Y,K}.

To extract physical information from the theory, we 
formulate the theory in terms of 
invariant dynamical variables. It is well known that 
as a result of such a formulation the angular 
components of the scalar fields $(\bf n)$  are absorbed by the physical 
vector fields $V^p$ and $\psi^p$ in the unitary gauge 
 
\be \label{67} 
B^p=V_i^p\hat Y_{ij}V_j^p\,;\,F^p=\bar\psi_{\alpha}^p\hat X_{\alpha\beta} 
\psi_{\beta}^p, 
\ee 
where $\hat Y,\;\hat X$ are the ordinary 
matrices of vector meson and fermion mass couplings in the WS 
theory multiplied by the 
rescaling parameters $\chi^2$ and $\chi$, respectively. 
 
In the first order formalism, the 
action (\ref{model}) in terms of the Lichnerowicz variables has 
the form 
\be 
\label{30} 
W^E_{CUT}=[P_f, f; P_g, g^c, \bar P_{\vh}, \vh_c|t]= 
\int\limits_{t_1}^{t_2}dt\int 
d^3x\left[\sum\limits_{f=g, V, \psi}P_fD_0f-\bar P_{\vh}D_0\vh_c-N_c 
{\cal H}\right], 
\ee 
where 
\be 
\label{30a} 
{\cal H}=-\frac{\bar P_{\vh}^2}{4}+6\frac{P_g^2}{\vh_c^2}-{\vh_c^2} 
\bar B+\vh_c F^p+ {\cal H}_0^{SM}+\bar\lambda \vh_c^4 
\ee 
is the Hamiltonian density, $\bar B$ is a contribution of the potential part 
of bosonic fields ($g_c$, $V^p$) 
\be 
\label{31} 
\bar B=B^p-{1\over 6}({}^{(3)}R(g^c_{ij})+8\vh_c^{-1/2}\Delta\vh_c^{1/2})\, 
;~~~~~ \Delta\vh_c=\ik_i(g_c^{ij}\ik_j\vh_c), 
\ee 
$B$ and $F^p$ is given by (\ref{67}) 
and $\bar\lambda=\chi^4\lambda$;~~ 
$P_f, \bar P_{\vh} $ are the canonical momenta of the corresponding fields, 
for example 
\be 
\label{33} 
D_0\vh_c=\ik_0\vh_c-\ik_k(N^k\vh_c) 
+{2\over 3}\vh_c\ik_kN^k\,,\,D_0g^c_{ij}= 
\ik_0g^c_{ij}-\nabla_iN_j-\nabla_jN_i+{2\over 3}\ik_k g^c_{ij}N^k . 
\ee 
These covariant derivatives multiplied by the factor $dt$ are invariant under 
kinemetric transformations \cite{YAF} 
\be 
\label{34} 
t\,\rightarrow\, t'=t'(t)\,;\,x^k\,\rightarrow \,x'^k=x'^k(t, x^1, x^2, 
x^3) \,,\, N\,\rightarrow\, N'... 
\ee 
This invariance means that GR and CUT represent an extended systems (ES) 
with constraints  and ``superfluous'' variables \cite{Dirac,KPP,G}. 
To separate  the physical sector of invariant variables and observables 
 from the parameters of general coordinate transformations, one 
needs the procedure of the Hamiltonian reduction, which leads to an equivalent 
unconstraint system, where one of ``superfluous'' variables becomes the 
dynamical parameter of evolution \cite{KPP,G}. 
 
The Hamiltonian reduction requires the evolution parameter of 
reduced system to be point out 
as one of the initial (superfluous) variables of the extended system. 
Such an evolution parameter in GR can be 
the global 
homogeneous component of the scale space factor \cite{Y,K,YAF,G} with 
a negative sign of its kinetic term. 
 
In our theory the role of the scale space factor is played by the scalar (Higgs) 
 field 
$\vh_c=\vh_{H_c}/\chi$ (see (\ref{9}) and (\ref{67a})). Therefore, we extract the evolution parameter 
by splitting the Higgs field and lapse fuction into two factors: 
homogeneous (global) and local 
\be 
\label{36} \bar \vh_c(x, t)=\vh_0(t)a(x, t);~~~~~~ 
N_c(x, t)=N_0(t){\cal N}(x, t); 
\ee 
the second factor $a(t,x)$, by definition, is constrained by 
the relation 
\be 
\label{37} 
\int d^3xa(x, t)\frac{D_0a(x, t)}{N_c}=0, 
\ee 
which diagonalizes the kinetic term of the action (\ref{30}). 
To get the conventional canonical structure for the new variables 
\be 
\label{38} 
\int d^3x(\bar P_\vh D_0\bar \vh_c)=\dot \vh_0\int d^3x\bar P_\vh a 
+\vh_0\int d^3x\bar P_\vh D_0 a 
=\dot \vh_0 P_0+\int d^3xP_aD_0a, 
\ee 
we define decomposition of $\bar P_\vh$ over the new momenta $P_0$ and 
$P_a$ conjugated to the new variables (\ref{36}) 
\be 
\label{40} 
\bar P_\vh=\frac{P_a}{\vh_0}+ 
P_0\frac{a}{{\cal N}V_0 }\,; 
~~~~~~~(\int d^3xa(x, t)P_a\equiv0,~~~V_0=\int d^3x\frac{a^2}{\cal N}). 
\ee 
The substitution of (\ref{40}) into the Hamiltonian part of the action 
(\ref{30}) extracts the ``superfluous'' momentum term 
\be 
\label{41} 
\int d^3xN_c{\cal H}=N_0\left[-\frac{P_0^2}{4V_0}+H_f\right]. 
\ee 
Finally, the extended action (\ref{30}) is 
\be 
W^E[P_f, f;P_0, \vh_0|t]= \int\limits_{t_1}^{t_2}dt\left(\left[\int 
d^3x \sum\limits_{f}P_fD_0f \right] -\dot \vh_0P_0 
-N_0\left[-\frac{P_0^2}{4V_0}+H_f\right]\right). 
\ee 
 
\section{Reduction and Dynamics of Proper Time} 
 
To remove arbitrariness connected with invariance of the theory 
with respect to time reparametrizations, we use the method 
of Hamiltonian reduction  ~\cite{KPP,G} where 
one of the dynamical variables transforms into the evolution parameter.

The reduction means explicit resolving of the constraint 
\be \label{45} 
\int d^3xN_c\frac{\delta 
W}{\delta N_c}=0\,\Rightarrow\, \frac{P_0^2}{4V_0}= H_f\equiv V_0 
\rho_{CUT}(\vh_0) 
\ee 
with respect to the momentum $P_0$. 
This equation has 
two solutions which correspond to two reduced systems with the actions 
\be 
\label{d7} 
W^R_{\pm}(P_f, f| 
\varphi_0)=\int\limits_{\varphi_1=\varphi_0(t_1)}^{\varphi_2=\varphi_0(t_2)} 
d\varphi_0 \left\{\left(\int 
d^3x\sum\limits_fP_fD_{\varphi}f\right)\mp2\sqrt{V_0 H_f}\right\} 
\ee 
where $\vh_0$ plays the role of the evolution parameter, and 
$D_\vh f={D_0f}/{\dot \vh_0}$ 
is the covariant derivative with the new shift vector $N^k$ and vector 
field $V$, which differs from the old ones by the factor 
$(\dot \vh_0)^{-1}$. 
 
The local equations of motion of the systems (\ref{d7}) reproduce 
the invariant sector of the initial extended system and determine 
the evolution of all variables $(P_f, f)$ with respect to the parameter 
$\vh_0$ 
\be 
\label{48} 
(P_f(x, t), f(x, t),\dots)\,\rightarrow\,(P_f(x, \vh_0), f(x, \vh_0),\dots). 
\ee 
The lapse function $N_0(t)$ forms a measurable time of an observer 
\be \label{50} 
 ds_c(dx=0)={\cal N}(x,t) N_0dt={\cal N}(x,\eta)d\eta; 
~~~~~~~~~~\,(\eta(t')=\eta(t)) 
\ee 
We call quantity $(\eta)$  the global conformal time. 
The functional $V_0$ in eq. (\ref{40}) can be 
chosen so that ${\cal N}(x, t)$ and ${a}(x,t)$ 
in the Newton approximation have the form 
\be 
\label{52} 
{\cal N}(x, t)=1+\delta_N(x)+\dots;~~~~~~~ 
a(x, t)=1+\delta_a(x)+\dots 
\ee 
where $\delta_N(x)$, $\delta_a(x)$ are the potentials 
of the Newton gravity. 
 
The reduced action (\ref{d7}) 
is completed by the equations of global dynamics: 
\be 
\label{d54} 
\frac{\delta W^E}{\delta N_0}=0\,\Rightarrow\, 
(P_0)_{\pm}=\pm2V_0\sqrt{\rho_{CUT}({\varphi_0})}; 
~~~~(\rho_{CUT}=\frac{H_f}{V_0}) 
\ee 
\be 
\label{d55} 
\frac{\delta W^E}{\delta {\varphi_0}}=0 
\,\Rightarrow\, P'_0=V_0\frac{d}{d{\varphi_0}}{\rho}_{CUT}({\varphi_0}) 
;~~~~~~~~(f'=\frac{d}{d\eta}f) 
\ee 
\be 
\label{d56} 
\frac{\delta W^E}{\delta P_0}=0\,\Rightarrow\, 
\left(\frac{d\varphi_0}{d\eta}\right)_{\pm} 
=\frac{(P_0)_{\pm}}{2V_0} 
=\pm\sqrt{\rho_{CUT}({\varphi_0})} 
\ee 
where the effective Hamiltonian density functional can be decomposed 
over powers of $(\varphi_0)$ 
\be \label{71} 
{\rho}_{CUT}= \frac{{\bf k}_A^2}{{\varphi_0}^2}+{\bf 
h}_R^2+{\bf\mu}_F^2{\varphi_0} 
+{\bf\Gamma}_B^{-2}{\varphi_0}^2+{\bf\Lambda}{\varphi_0}^4, 
\ee 
where the coefficients of the decomposition are the functionals of the 
local fields. 
 
\section{Cosmic Higgs vacuum} 
 
Equations (\ref{d54}), (\ref{d55}), and (\ref{d56}) lead 
to the Friedmann-like evolution of global conformal time of an 
observer 
\be 
\label{70} 
\eta({\varphi_0})=\int\limits_0^{\varphi_0}d\varphi 
{\rho}^{-1/2}_{CUT}(\varphi), 
\ee 
and to the conservation law 
\be 
\label{72} 
\frac{({\bf k}_A^2)'}{{\varphi_0}^2}+ 
({\bf h}_R^2)'+({\bf\mu}_F^2)'{\varphi_0}+ 
({\bf\Gamma}_B^{-2})'{\varphi_0}^2+({\bf\Lambda})'{\varphi_0}^4=0. 
\ee 
The red shift and the Hubble law in the conformal time version 
\be 
\label{73} 
z(D_c)=\frac{{\varphi_0}(\eta_0)}{{\varphi_0}(\eta_0-D_c)}-1\simeq 
D_cH_{Hub};~~~~~~~~~~\,H_{Hub} 
=\frac{1}{{\varphi_0}(\eta)}\frac{d}{d\eta}{\varphi_0}(\eta) 
\ee 
reflect the alteration of the size of atoms in the process of evolution of 
masses ~\cite{N,KPP}. 
 
In the dependence on the value of ${\varphi_0}$, there is dominance of the 
kinetic or the potential part of the Hamiltonian (\ref{71}), (\ref{72}), and 
different stages of evolution of the Universe (\ref{70}) can appear: 
anisotropic $({\bf k}_A^2\ne0)$ and radiation $({\bf h}_R^2\ne0)$ (at the 
beginning of the Universe), dust $({\bf\mu}_F^2\ne\,;\,{\bf\Gamma}_B^{-2})$ 
and De-Sitter $\Lambda\ne0$ (at the present time). 
 
In perturbation theory, the factor $a(x,t)=(1+\delta_{a})$ 
represents the potential of the Newton gravity $(\delta_{a})$. 
Therefore, the Higgs-PCT 
field, in this model, has no particle-like excitations 
(as it was predicted in paper \cite{PR}).

For an observer, who lives in the Universe, 
a state of ``vacuum'' is the state of the Universe at the present time: 
$|Universe>=|Lab.vacuum>$, 
as his unified theory pretends to describe both observational cosmology and 
any laboratory experiments. 
 
In correspondence with this definition, the Hamiltomian $H_f$ 
can be  split into the large (cosmological -- global) and small (laboratory 
-- local) parts 
\be\label{d5} 
H_f[\varphi_0] 
\buildrel{\rm 
def}\over 
= 
\rho_0V_0+(H_f-\rho_0V_0)= 
\rho_0(\varphi_0)V_0+H_L 
\ee 
where the global  part of the 
Hamiltonian $\rho_0(\varphi_0)V_0$ can be defined as the ``Universe'' 
averaging so that the ``Universe'' averaging of the local part of 
the Hamiltonian (\ref{d5}) is equal to zero 
\be\label{d28} 
<Universe|H_f|Universe>=\rho_0V_0,~~~~~~ 
<Universe|H_L|Universe>=0. 
\ee 
 
Let us suppose that the local dynamics $(H_L)$ can be neglected if we 
consider the cosmological sector of the proper time dynamics (\ref{d54}), 
(\ref{d55}), (\ref{d56}). In this case,  eqs.(\ref{d56}), and (\ref{73}) 
give the relation 
between the present-day value of the scalar field and 
the cosmological observations 
\be 
\label{d12a} 
\bar\varphi(\eta=\eta_0)=\frac{\sqrt{\rho_0(\eta_0)}}{H_0(\eta_0)}. 
\ee

The present-day mean matter density 
\be\label{d30} 
\rho_b=\Omega_0\rho_{cr};~~~~~~~~~~~~~(\rho_{cr}= 
\frac{3H_0}{{8\pi}}M_{Pl}^2      ) 
\ee 
is estimated from experimental data on 
luminous matter ($\Omega_0=0.01$), the flat rotation curves of spiral 
galaxies ($\Omega_0=0.1$) and others data~\cite{rpp} ($0.1<\Omega_0<2$). 
 
We should also take into account that these observations 
reflects the density at the time of radiation of 
light from cosmic objects $\Omega(\eta_0-distance/c)$, 
which was less than at the present-day density $\Omega(\eta_0)=\Omega_0$ 
due to an increasing mass of the matter. 
This effect of retardation can be roughly estimated by the averaging 
of $\Omega(\eta_0-distance/c)$ over distances (or proper time) 
${\gamma}={\eta_0 \Omega_0}/{\int\limits_0^{\eta_0} d\eta\Omega(\eta)}$. 
For the dust stage the coefficient of the increase is $\gamma=3$. 
Finally, we get the relation of the cosmic 
value of the Planck ``constant'' and the GR one 
\be\label{d32} 
\frac{\bar\varphi(\eta=\eta_0)}{M_{Pl}}\sqrt{{8\pi\over 3}} 
=\sqrt{\gamma\Omega_{0(exp)}}/h=\omega_0, 
\ee 
where $h=0.4\div 1$ is the observational bounds for the Hubble parameter. 
 
From data on $\Omega_0$ we can estimate $\omega_0$: $\omega_0=0.04$ 
(luminous matter), $\omega_0=0.4$ (flat rotation curves of spiral 
galaxies), and $0.4<\omega_0<9$ (others data~\cite{rpp}) for lower values of 
$h$ ($h=0.4$). 
 
The second term of the decomposition of the reduced action (\ref{d7}) over 
$V_0^{-1}$ defines the action for local excitations 
\be 
P_0d\varphi_0=2V_0\sqrt{\rho_0(\bar\varphi_0)}d\bar\varphi_0+ 
H_L(\bar\varphi_0)d\eta 
+o\left(\frac{1}{V_0}\right) 
\ee 
in terms of the measurable time $\eta$. 
 
Really, an observer uses the action for description of 
laboratory 
experiments in a very small interval of time in comparison with the 
lifetime of the Universe $\eta_0$: 
$\eta_1=\eta_0-\xi\,;\,\eta_2=\eta_0+\xi\,;\,\xi\ll\eta_0,$ 
and during this time-interval  $\varphi_0(\eta)$  can be considered 
as constant 
$\varphi_0(\eta_0+\xi)\approx\varphi_0(\eta_0)=M_{Pl}\sqrt{{3}/{8\pi}}$. 
Thus, we got the $\sigma-$model version of the standard model ~\cite{PR}.

\section{Conclusion} 
 
The conformal unified theory (CUT) of strong, electroweak and gravitation 
interactions from a physical point of view indentifies the 
Higgs scalar field in SM with the determinant of the space metric (i.e. scale 
factor) in the theory of gravity. 
This indentification leads to important physical 
consequences: 
 
CUT doesn't  need the Higgs potential for the formation of the homogeneous 
part of the Higgs field. 
The homogeneous part is extracted by the Hamiltonian reduction as an evolution 
parameter of the reduced system. The  proper time of an observer becomes the 
dynamical variable with respect to the 
evolution parameter with the Friedmann-like cosmological 
equation. In contrast with the conventional Higgs effect (where 
the Higgs field is determined by  parameters of the vacuum state), 
the Higgs field in CUT is determined by the integrals of motion and initial 
data of the state of the Universe (its density $\rho$ and time of life 
$H_0^{-1}$):  $ \phi_0=\sqrt{\rho_0}/H_0, $ in satisfactory agreement with 
observational data.  
 
{\bf Acknowledgments} 
 
We are happy to acknowledge interesting and critical 
discussions with Profs. B.M. Barbashov, R. Brout, A.V. Efremov, 
G.A. Gogilidze, V.G. Kadyshevsky, 
A.M. Khvedelidze, W.Kummer,  and Yu.G. Palii. 
We also thank 
the Russian Foundation  for Basic 
Researches, Grant N 96\--01\--01223 
and the Polish 
Committee for Scientific Researches, Grant N 603/P03/96, for support.

\end{document}